\documentclass[twocolumn,showpacs,aps,floatfix]{revtex4}

\usepackage[latin1]{inputenc}
\usepackage{amsfonts}
\usepackage{amsmath}
\usepackage[usenames]{color}
\usepackage{amssymb}
\usepackage[american]{babel}
\usepackage{graphicx}
\usepackage{subfigure}
\usepackage{multirow}
\usepackage{psfrag}
\newcommand{\ket}[1]{|#1\rangle}
\newcommand{\bra}[1]{\langle#1|}

\begin{document}

% front matter
\title{Vacuum entanglement enhancement by a weak gravitational field}
\author{M.~Cliche and A.~Kempf}
\affiliation{Department of Applied Mathematics,
University of
Waterloo, Waterloo, Ontario, Canada N2L 3G1}

\date{\today}

\begin{abstract}
Separate regions in space are generally entangled, even in the vacuum state. It is known that this entanglement can be swapped to separated Unruh-DeWitt detectors, i.e., that the vacuum can serve as a source of entanglement. Here, we demonstrate that, in the presence of curvature, the amount of entanglement that Unruh-DeWitt detectors can extract from the vacuum can be increased.
\end{abstract}
 \maketitle

\section{Introduction}

Two Unruh-DeWitt detectors that interact with a quantum field in the vacuum state have access to a renewable source of entanglement \cite{Rezn1,Rezn2, cliche-kempf,Hu,Hu2}, namely by swapping entanglement from the quantum field. In this context, it was recently shown \cite{Meni} that, in an expanding space-time,
the entanglement of the vacuum decreases significantly due to the effects of the Gibbons-Hawking temperature \cite{Hawking}. While this example showed that gravity is able to act as a decohering agent, we will here show that gravity can also act to enhance entangelement-related phenomena. Namely, we will show that a weak gravitational field, such as that caused by a planet, can enhance the extraction of entanglement from the vacuum.

This article is organized as follows.  In Sec. \ref{vac_en} we review the extraction of entanglement with Unruh-DeWitt detectors in Minkowski space-time.  In Sec. \ref{Newton} we review the Newtonian limit of general relativity and in Sec. \ref{blue} we look at Unruh-DeWitt detectors in the presence of weak gravity. Then, in Sec. \ref{Prop} we compute the first order correction to the propagator on the perturbed background.  In Sec. \ref{negativity} we calculate explicitly the entanglement between the two detectors near a spherically symmetric star.  In the last section, we propose extensions.

We work with the natural units $\hbar=c=G=1$ and the Minkowski metric $\eta_{\mu\nu}=diag(-1,1,1,1)$.  We denote the coordinate time by $x^0=t$ while the proper time is denoted by $\tau$.  Wherever necessary to avoid ambiguity we will denote operators $O$ or states $\ket{\psi}$ which live in the Hilbert space ${\cal H}^{(j)}$ of the $j$'th subsystem by a superscript $(j)$, for example, $O^{(j)}$ and $\ket{\psi^{(j)}}$.  Orders in perturbation theory will be denoted by a subscript (j), as in, e.g., $P=P_{(0)}+P_{(1)}+O(\epsilon^2)$.  We work in the interaction picture.

\section{Vacuum entanglement}\label{vac_en}

Let us first briefly review vacuum entanglement with Unruh-DeWitt detectors \cite{Rezn1,Rezn2,Meni}.  To begin, let us denote the overall Hilbert space by ${\cal H}={\cal
H}^{(1)}\otimes{\cal H}^{(2)}\otimes{\cal H}^{(3)}$, where the first two
Hilbert spaces belong to two Unruh-DeWitt detectors and where
the third Hilbert space is that of a quantum massless scalar field. The total Hamiltonian $H$ of the system with respect to the coordinate time $t$ is
\begin{eqnarray}
H&=&H_F+H_D+H_{int}\nonumber\\
H_F&=&\frac{1}{2} \int d^3x \left[\pi^2(x)+\left(\nabla\phi(x)\right)^2 \right]\nonumber\\
H_D&=&\sum_{j=1}^{2} \Big[(E_g+\Delta E)\ket{e^{(j)}}\bra{e^{(j)}}\nonumber\\
&&+E_g\ket{g^{(j)}}\bra{g^{(j)}}\Big]\frac{d\tau_j(t)}{d t}\nonumber\\
H_{int}(t)&=&\sum_{j=1}^{2}\alpha_j\eta(\tau_j(t))
\Big(\ket{e^{(j)}}\bra{g^{(j)}}e^{i\Delta E \tau_j(t)}\nonumber\\
&&+\ket{g^{(j)}} \bra{e^{(j)}}e^{-i\Delta E \tau_j(t)}\Big)\phi\left(x_j\left(\tau_j(t)\right)\right)\frac{d\tau_j(t)}{d t}\nonumber\\
\label{ty}
\end{eqnarray}
where $H_F$ is the Hamiltonian of a free massless scalar field, $H_D$ is the Hamiltonian of the two detectors, $H_{int}(t)$ is the interaction Hamiltonian \cite{Bir} in the interaction picture, $\alpha_j$ is the coupling constant of the $j$'th detector ($j\in\{1,2\}$), $\phi(x_j(\tau_j))$ is the field at the point of the $j$th detector and
$m^{(j)}(\tau_j):=\left(\ket{e^{(j)}}\bra{g^{(j)}}e^{i\Delta E \tau_j}+\ket{g^{(j)}}\bra{e^{(j)}}e^{-i\Delta E \tau_j}\right)$
is the monopole matrix of the $j$th detector. The function $\eta(\tau_j)$ will be
used to describe the continuous switching on and off of the detectors and $\tau_j$ is the proper time of the $j$th detector.

Let us first consider the special case where $\tau_1(t)=\tau_2(t)=\tau(t)$ such that the evolution operator $U=Te^{-i\int dt H_{int}(t)}$ acting on states takes the form:
\begin{eqnarray}
U&=&T \mathrm{exp}\Bigg\{-i\int d\tau \Bigg[ \alpha_1 \eta(\tau)m^{(1)}(\tau)\phi\left({x}_1(\tau)\right)\nonumber\\
&&+  \alpha_2\eta\left(\tau\right)m^{(2)}\left(\tau\right)\phi\left({x}_2\left(\tau\right)\right)\Bigg]\Bigg\}.
\end{eqnarray}
We assume that the initial state of the system is $\ket{0g^{(1)}g^{(2)}}$. After the unitary evolution of the total system, we trace out the field and obtain at $O(\alpha^2)$ \cite{Rezn1,Meni}:
\begin{eqnarray}
\rho_f^{(1,2)}&=&Tr_{(3)}\Big(U \ket{0g^{(1)}g^{(2)}}\bra{0g^{(1)}g^{(2)}}U^{\dag}\Big)\nonumber\\
&=&\left(\begin{smallmatrix} 0&0&0& X\\ 0& P_{1}&  Y&0 \\ 0 &  Y^{\ast} &  P_{2} & 0 \\  X^{\ast}&0&0&1-P_{1}-P_{2}
\end{smallmatrix} \right)+O(\alpha^4)
\end{eqnarray}
in the basis $\ket{e^{(1)}e^{(2)}}$, $\ket{e^{(1)}g^{(2)}}$, $\ket{g^{(1)}e^{(2)}}$ and $\ket{g^{(1)}g^{(2)}}$. The matrix elements $P_j$, $X$ and $Y$ read
\begin{eqnarray}
P_{j}&=& \alpha_j^2\int_{-\infty}^{\infty}d\tau\int_{-\infty}^{\infty}d\tau ' \eta(\tau)\eta(\tau ') e^{-i\Delta E (\tau-\tau ')} \nonumber\\
&&\times D\left({x}_j(\tau),{x}_j(\tau ')\right) \label{Pj}\\
X&=&-\alpha_1\alpha_2\int_{-\infty}^{\infty}d\tau\int_{-\infty}^{\tau}d\tau ' \eta(\tau)\eta(\tau ') e^{i\Delta E(\tau+\tau ')} \nonumber\\
&&\times\left(D\left({x}_1(\tau),{x}_2(\tau ')\right)+D\left({x}_2(\tau),{x}_1(\tau ')\right) \right) \label{Xj}\\
Y&=&  \alpha_1\alpha_2\int_{-\infty}^{\infty}d\tau\int_{-\infty}^{\infty}d\tau ' \eta(\tau)\eta(\tau ') e^{-i\Delta E(\tau-\tau ')}\nonumber\\
&&\times D\left({x}_1(\tau),{x}_2(\tau ')\right)
\end{eqnarray}
where $D(x,y)=\bra{0}\phi(x)\phi(y)\ket{0}$. To measure the entanglement of $\rho_f^{(1,2)}$, we use the negativity \cite{Vid} which gives:
\begin{eqnarray}
N&=&\max\Big(\sqrt{(P_{1}-P_{2})^2+4|X|^2}-P_{1}-P_{2},0\Big)\nonumber\\
&&+O(\alpha^4).\label{N}
\end{eqnarray}
In order to obtain more explicit results, let us consider, for example, the case where the detectors are inertial and separated by a constant proper distance $L_p$ in Minkowski space-time, in which case $x_j^0(\tau)=\tau$.  Let us also assume that $\alpha_1=\alpha_2=\alpha$ such that in Minkowski space we have $P_{1}=P_{2}=P$ and $N=2\max(|X|-P)+O(\alpha^4)$. For simplicity we choose the switching functions to be gaussian: $\eta(\tau)=e^{-\tau^2/(2\sigma^2)}$.

We will call $X$ the exchange term and $P_{j}$ the local noise term.  This is because $X$ can be interpreted as describing the exchange of virtual quanta between the two detectors, and $P_j$ can be interpreted as describing the detection of virtual quanta by detector $j$.   In order to allow the introduction of gravity (which will enter mostly through the propagator), it will be useful to view $P_j$ and $X$ as functions of the propagator.  To this end, let us already ensure that the time ordering is respected.  For $X$ this is straightforward since the time integrations respect time ordering by construction.  We can still simplify $X$ by using the variable change $s=\tau-\tau '$ and $u=\tau+\tau '$ such that we have
\begin{eqnarray}
X&=& -\alpha^2 e^{-\sigma^2\Delta E^2}\sigma \sqrt{\pi} \int_{0}^{\infty}ds e^{-s^2/(4\sigma^2)} \nonumber\\
&&\times\left(G(\vec{x}_1,\vec{x}_2,s)+G(\vec{x}_2,\vec{x}_1,s)\right) \label{X}
\end{eqnarray}
where $G(x,y)=\bra{0}T\phi(x)\phi(y)\ket{0}=G(\vec{x},\vec{y},x^{0}-y^{0})$ is Feynman propagator \cite{Pesk}.  For $P_j$, we introduce a convenient change of variables for the
double integral over the ($\tau,\tau '$) plane \cite{Satz07}, making $u=\tau$, $s=\tau-\tau '$ in the lower half-plane $\tau '<\tau$ and $u=\tau '$, $s=\tau '-\tau$ in the upper half-plane $\tau<\tau '$, $P_j$ becomes:
\begin{eqnarray}
P_{j}&=&2\alpha^2\sigma \sqrt{\pi} \Re \left(\int_{0}^{\infty}ds e^{-s^2/(4\sigma^2)-i\Delta E s} G(\vec{x}_j,\vec{x}_j,s) \right).\nonumber\\ \label{Pe}
\end{eqnarray}
In Minkowski space-time we use the Boulware vacuum such that the propagator is given by
\begin{eqnarray}
G(x,y)=\frac{-1}{4\pi^2\left[(x^0-y^{0})^2-|\vec{x}-\vec{y}|^2-i\epsilon \right]}\label{propagator}.
\end{eqnarray}
where $\lim_{\epsilon\rightarrow 0^+}$ is implicit.  Using Eq. (\ref{propagator}) in Eq. (\ref{Pe}) and (\ref{X}) we obtain the local noise and the exchange term in Minkowski space-time \cite{Meni},
\begin{eqnarray}
P&=&\frac{\alpha^2}{4\pi} \left(e^{-\Delta E^2\sigma^2}-\Delta E\sqrt{\pi}\sigma \mathrm{erfc}(\Delta E \sigma) \right)\label{PG}\\
X&=&\frac{\alpha^2\sigma i}{4L_p\sqrt{\pi}} e^{-\Delta E^2\sigma^2-L_p^2/4\sigma^2} \mathrm{erfc}\left(\frac{-iL_p}{2\sigma}\right)\label{XG}
\end{eqnarray}
where $L_p=|\vec{x}_1-\vec{x}_2|$ and $\mathrm{erfc}(x)=1-\mathrm{erf}(x)$. In the regime $\Delta E \sigma \rightarrow \infty$ and $L_p/\sigma\rightarrow \infty$, we have $P\approx \frac{\alpha^2 e^{-\Delta E^2\sigma^2}}{8\pi\Delta E^2\sigma^2}$ and $
|X|\approx \frac{\alpha^2 \sigma^2 e^{-\Delta E^2\sigma^2}}{2\pi L_p^2}$.  Therefore, to get a non vanishing negativity, we need $\Delta E \sigma > \frac{L_p}{2\sigma}$.  To optimize the negativity in that regime, we set $\frac{\partial N}{\partial \Delta E_{opt}}=0$, which yields $\Delta E_{opt}\approx \frac{L_p}{2\sigma^2}\left(1+2\sigma^2/L_p^2\right)$.  The resulting negativity is then $N_{opt}\approx\frac{4\alpha^2\sigma^4 e^{-L_p^2/{4\sigma^2}}}{\pi L_p^4}$.

\section{Newtonian limit} \label{Newton}

Let us now briefly review the Newtonian limit of general relativity, see e.g. \cite{schutz}.  In this limit we can write the metric as $g_{\alpha\beta}=\eta_{\alpha\beta}+h_{\alpha\beta}$ where $|h_{\alpha\beta}|\ll 1$.  Note that under a small change of coordinates $x^{\mu}\rightarrow x^{\mu}+\xi^{\mu}$ the term $h_{\alpha\beta}$ has a gauge transformation $h_{\alpha\beta}\rightarrow h_{\alpha\beta}-\partial_{\beta}\xi_{\alpha}-\partial_{\alpha}\xi_{\beta}$.  Let us define the quantity $\bar{h}^{\mu\nu}:=h^{\mu\nu}-\eta^{\mu\nu}h^{\alpha}_{\alpha}/2$.  To simplify the Einstein equation, we choose to work in the Lorentz gauge in which ${\bar{h}^{\mu\nu},}_{\nu}=0$.  In this gauge, the linearized Einstein equation reads $\partial_{\alpha}\partial^{\alpha} \bar{h}^{\mu\nu}=-16\pi T^{\mu\nu}$.  In the Newtonian limit the gravitational field is too weak to produce velocities near the speed of light, thus only the $T^{00}$ component of the stress-energy tensor contributes significantly and we can make the approximation $\partial_{\alpha}\partial^{\alpha}\approx \nabla^2$.  This means that the Einstein equation can be approximated as $\partial_{\alpha}\partial^{\alpha}\bar{h}^{00}\approx\nabla^2\bar{h}^{00}\approx-16\pi\rho$.  From this we conclude that the dominant component of $\bar{h}^{\mu\nu}$ is $\bar{h}^{00}$, such that in terms of $h^{\alpha\beta}$ we have $h^{00}=h^{ii}=\bar{h}^{00}/2$.  Thus, the line element takes the form:
\begin{eqnarray}
ds^2&=&-\left(1-\bar{h}^{00}/2\right)dt^2\nonumber\\
&&+\left(1+\bar{h}^{00}/2\right)\left(dx^2+dy^2+dz^2\right).\label{metric}
\end{eqnarray}

Now assume we have a compact object, say a star of dark matter that does not interact with the quantum field  and is of radius $R_o$ and of constant density $\rho=3M/(4\pi R_o^3)$. We solve $\nabla^2\bar{h}^{00}\approx-16\pi\rho$ with the usual boundary conditions $\bar{h}^{00}(|\vec{r}|\rightarrow\infty)=0$, $\frac{\partial\bar{h}^{00}(\vec{r})}{\partial r}(r=0)=0$ and with the continuity conditions $\bar{h}^{00}(|\vec{r}|\rightarrow R_o-\epsilon)=\bar{h}^{00}(|\vec{r}|\rightarrow R_o+\epsilon)$, $\frac{\partial \bar{h}^{00}}{\partial r}(|\vec{r}|\rightarrow R_o-\epsilon)=\frac{\partial \bar{h}^{00}}{\partial r}(|\vec{r}|\rightarrow R_o+\epsilon)$ in the limit $\epsilon\rightarrow 0$.  This gives
\begin{eqnarray}
\bar{h}^{00}(\vec{r})&=&\displaystyle \left\{ \begin {array}{ll} \frac{2M}{R_o}\left(3-\frac{|\vec{r}|^2}{R_o^2} \right) & \mbox{when}\,\, |\vec{r}| < R_o, \\
\frac{4M}{|\vec{r}|}  & \mbox{when}\,\, |\vec{r}| > R_o\\
\end{array} \right. \label{hb}
\end{eqnarray}
so to have $|h_{\alpha\beta}|\ll 1$ we require $M/R_o\ll 1$.

\section{Detectors on the curved background}\label{blue}

Let us now consider the two Unruh-DeWitt detectors on the background of the weak gravitational field.  We assume that the two detectors and the center of the star are all on a same axis.  Therefore, detector 1 is located at a fixed distance $r_1$ from the center of the star and similarly detector 2 is located at $r_2=r_1+L$ from the center of the star.  This means that their proper times do not coincide $\tau_1(t)\neq\tau_2(t)$, so we may write the evolution operator as
\begin{eqnarray}
U&=&T \mathrm{exp}\Bigg\{-i\int d\tau_1 \alpha\Bigg[ \eta(\tau_1)m^{(1)}(\tau_1)\phi\left({x}_1(\tau_1)\right)\nonumber\\
&&+  \eta\left(\tau_2(\tau_1)\right)m^{(2)}\left(\tau_2(\tau_1)\right)\phi\left({x}_2\left(\tau_2(\tau_1)\right)\right)\frac{d\tau_2(\tau_1)}{d\tau_1}\Bigg]\Bigg\}\nonumber\\
\end{eqnarray}
and using Eq. (\ref{metric}) we have:
\begin{eqnarray}
\tau_2(\tau_1)&=&\tau_1\sqrt{\frac{1-\bar{h}^{00}(r_2)/2}{1-\bar{h}^{00}(r_1)/2}}\nonumber\\
&=&\tau_1\left(1-\frac{\bar{h}^{00}(r_2)}{4}+\frac{\bar{h}^{00}(r_1)}{4}+O\left([\bar{h}^{00}]^2\right)\right). \nonumber\\
\end{eqnarray}
To simplify our analysis we want to avoid this blueshift effect. To do this, we assume that the two detectors are close enough such that their internal clocks have the same speed at first order in perturbation theory.  This will be so if $|\bar{h}^{00}(r_2)/4-\bar{h}^{00}(r_1)/4|\lesssim O\left([\bar{h}^{00}(r_2)]^2\right)$ which for detectors outside the star gives $L\lesssim 16 M $.  Under that assumption, we have $\tau_2=\tau_1\left(1+O\left([\bar{h}^{00}]^2\right)\right)$ such that one can easily verify that Eq. (\ref{Pj}) and Eq. (\ref{Xj}) still hold up to $O\left([\bar{h}^{00}]^2\right)$.

We are therefore left with two first order contributions to the exchange term $X$ and the local noise term $P_{j}$, the first one which we denote by $\tilde{X}_{(1)}$ and $\tilde{P}_{j(1)}$ is essentially a result of the time dilation caused by the star and the second one which we denote by $\delta X_{(1)}$ and $\delta P_{j(1)}$ comes from a modification of the propagator on the curved background. Let us denote the perturbative expansion of the propagator as $G(x,y)=G_{(0)}(x,y)+G_{(1)}(x,y)$ and since it is widely believed that a Boulware-like vacuum is the right vacuum for a quantum field in a Newtonian gravitational potential \cite{Satz08,Satz05}, we use Eq. (\ref{propagator}) for $G_{(0)}(x,y)$.  We can easily evaluate the contributions $\tilde{X}_{(1)}$ and $\tilde{P}_{j(1)}$ by first noting that
\begin{eqnarray}
x_j(\tau_j)=\left(\frac{\tau_j}{\sqrt{1-\bar{h}^{00}(r_j)/2}},\vec{r}_j\right).
\end{eqnarray}
Thus, when $L\lesssim 16 M$ we have using Eq. (\ref{propagator})
\begin{eqnarray}
&&G_{(0)}\left(|\vec{x}_1(\tau)-\vec{x}_2(\tau ')|,{x}^0_1(\tau)-{x}^0_2(\tau ')\right)=\nonumber\\
&&\left(1-\bar{h}^{00}(r_1)/2\right)G_{(0)}\left( L_p\left(1-\bar{h}^{00}(r_1)/2\right),\tau-\tau '\right)\nonumber\\
&&+O\left([\bar{h}^{00}]^2\right)
\end{eqnarray}
where
\begin{eqnarray}
L_p&:=&\int_{r_1}^{r_1+L}\sqrt{1+\bar{h}^{00}(r)/2}dr\nonumber\\
&\approx& L(1+\bar{h}^{00}(r_1)/4)
\end{eqnarray}
is the proper distance between the two detectors.  Hence, when we put this back in Eq. (\ref{Pe}) and Eq. (\ref{X}) we have the first order corrections
\begin{eqnarray}
\tilde{P}_{j(1)}&=&-\frac{\bar{h}^{00}(r_j)}{2} P_{(0)} \label{TP}\\
\tilde{X}_{(1)}&=&-\frac{\bar{h}^{00}(r_1)}{2} \left(X_{(0)}+L_p\frac{\partial X_{(0)}}{\partial L_p} \right)\label{TX}
\end{eqnarray}
where the zeroth order terms are given by Eq. (\ref{PG}) and (\ref{XG}).

\section{Correction to the propagator}\label{Prop}

In this section we compute the first order correction to the propagator on the perturbed background. The first steps of our calculation can be found in \cite{Satz08}. To focus on the correction caused by gravity, we assume that the field is minimally coupled to curvature and to the matter that composes the star.  Under these assumptions, the propagator is a Green's function of the Klein-Gordon operator
\begin{eqnarray}
\Box_{x} G(x,y)=\frac{i\delta(x,y)}{\sqrt{-g(x)}}
\end{eqnarray}
where $\Box_x f(x)=\frac{1}{\sqrt{-g}}\partial_{\mu}\left[ \sqrt{-g}g^{\mu\nu}\partial_{\nu}f(x)\right]$.  The first order correction to $g$ is $g=-1-h^{\alpha}_{\alpha}=-1-\bar{h}^{00}$.  Using $G(x,y)=G_{(0)}(x,y)+G_{(1)}(x,y)$ we have:
\begin{eqnarray}
&&\frac{1}{\sqrt{1+h^{\alpha}_{\alpha}}}\partial_{\mu}\Big[\sqrt{1+h^{\alpha}_{\alpha}} \left(\eta^{\mu\nu}-h^{\mu\nu} \right) \nonumber\\
&&\times\partial_{\nu} \left(G_{(0)}(x,y)+G_{(1)}(x,y) \right)\Big] = \frac{i\delta(x,y)}{\sqrt{1+h^{\alpha}_{\alpha}}}.
\end{eqnarray}
Expanding everything to first order only and using the fact that $G_{(0)}(x,y)$ solves the zeroth order equation, we obtain
\begin{eqnarray}
-h^{\mu\nu}\partial_{\mu}\partial_{\nu}G_{(0)}(x,y)&+&\Box_{(0)x}G_{(1)}(x,y)\nonumber\\
-\partial_{\mu}h^{\mu\nu}\partial_{\nu}G_{(0)}(x,y)&+&\partial_{\mu}(h^{\alpha}_{\alpha}/2)\eta^{\mu\nu}\partial_{\nu}G_{(0)}(x,y)\nonumber\\
&=&-i\delta(x,y)h^{\alpha}_{\alpha}/2
\end{eqnarray}
where we used $\Box_{(0)x}:=\eta^{\mu\nu}\partial_{\mu}\partial_{\nu}$.  Using again the fact that $i\delta(x,y)=\Box_{(0)x} G_{(0)}(x,y)$ we can simplify the previous equation,
\begin{eqnarray}
\Box_{(0)x} G_{(1)}(x,y)&=&\partial_{\mu}\bar{h}^{\mu\nu}\partial_{\nu}G_{(0)}(x,y)\nonumber\\
&&+\bar{h}^{\mu\nu}\partial_{\mu}\partial_{\nu}G_{(0)}(x,y)\nonumber\\
&=&\bar{h}^{\mu\nu}\partial_{\mu}\partial_{\nu}G_{(0)}(x,y)\nonumber\\
&=&\bar{h}^{00}(x)\partial^2_{x^0}G_{(0)}(x,y)\label{prop1a}
\end{eqnarray}
where we used the fact that we are in the Lorentz gauge and that in the Newtonian limit $\bar{h}^{00}$ is the dominant component of $\bar{h}^{\mu\nu}$.  Note that since the space-time we consider is static and asymptotically flat, the propagator $G(x,y)$ can be seen as the analytic continuation of the unique Green's function on the positive definite section \cite{Satz08}.  Since this holds order by order in perturbation theory, at first order perturbation we can use $G_{(0)}$ as the inverse of $\Box_{(0)}$ such that:
\begin{eqnarray}
G_{(1)}(x,y) = -i \int d^4z G_{(0)}(x,z)\bar{h}^{00}(z)\partial^2_{z^0}G_{(0)}(z,y). \label{prop1b}
\end{eqnarray}
This equation gives us explicitly the first order correction to the propagator.  It is clear from this equation that the entire space-time perturbation will modify the propagator, and the most significant contribution will come from the patch of space-time near $x$ and $y$.  We now insert $G_{(0)}(x,y)$ in Eq. (\ref{prop1b}) and using the fact that $\bar{h}^{00}(x)$ is independent of time, we obtain
\begin{eqnarray}
G_{(1)}(x,y)&=&\frac{-i}{16\pi^4} \int dz^0d^3z \bar{h}^{00}(\vec{z}) \nonumber\\
&&\times\Bigg[\frac{8(\tilde{z}+s)^2}{(\tilde{z}-z_1)^3(\tilde{z}-z_2)^3(\tilde{z}-z_o)(\tilde{z}+z_o)}\nonumber\\
&&-\frac{2}{(\tilde{z}-z_1)^2(\tilde{z}-z_2)^2(\tilde{z}-z_o)(\tilde{z}+z_o)}  \Bigg]\nonumber\\
\end{eqnarray}
where we use the definitions $Z_x:=|\vec{x}-\vec{z}|$, $Z_y:=|\vec{y}-\vec{z}|$, $s:=x^0-y^0$, $\tilde{z}:=z^0-x^0$, $z_o:=X+i\epsilon$, $z_1:=-s+Y+i\epsilon$ and $z_2:=-s-Y-i\epsilon$.  We can then perform the $z^0$ integration with the residue theorem.  We choose a closed contour in the upper half of the complex plane and the upper part of the contour is equal to zero because the integrand vanishes sufficiently rapidly as $z^0=R e^{i\theta}\Big|_{R\rightarrow \infty}$.  We thus have:
\begin{eqnarray}
G_{(1)}(x,y)&=&\frac{1}{8\pi^3} \int d^3z \bar{h}^{00}(\vec{z}) \Bigg[\nonumber\\
&&\frac{8(z_o+s)^2}{(z_o-z_1)^3(z_o-z_2)^3 2z_o} \nonumber\\
&&+4\frac{d^2}{d\tilde{z}^2}\left(\frac{(\tilde{z}+s)^2}{(\tilde{z}-z_2)^2(\tilde{z}-z_o)(\tilde{z}+z_o)}\right)\Big|_{\tilde{z}=z_1} \nonumber\\
&&-\frac{2}{(z_o-z_1)^2(z_o-z_2)^2 2z_o} \nonumber\\
&&-\frac{d}{d\tilde{z}}\left(\frac{2}{(\tilde{z}-z_2)^2(\tilde{z}-z_o)(\tilde{z}+z_o)}\right)\Big|_{\tilde{z}=z_1}  \Bigg]\nonumber\\
&=&\frac{1}{8\pi^3}\int d^3z \bar{h}^{00}(|\vec{z}|) \nonumber\\
&&\times\left[\frac{3(s^2+Z_xZ_y)(Z_x+Z_y)+Z_x^3+Z_y^3}{(Z_xZ_y+i\epsilon)(s^2-[Z_x+Z_y+i\epsilon]^2)^3}  \right].\nonumber\\\label{prop1}
\end{eqnarray}
Using the above equation we may now evaluate $\delta P_{j(1)}$ and $\delta X_{(1)}$.  For $\delta P_{j(1)}$, we have $Z_x=Z_y=Z$, such that the correction to the propagator can be greatly simplified with a simple change of variables
\begin{eqnarray}
G_{(1)}(\vec{x},\vec{x},s)&=&\frac{1}{2r\pi^2}\int_0^{\infty} dR R \bar{h}^{00}(R) \int_{|r-R|}^{r+R}dv \nonumber\\
&&\times\frac{3s^2+4v^2}{(s^2-4v^2-i\epsilon)^3} \label{G1}
\end{eqnarray}
where $r$ is the distance between $\vec{x}$ and the center of the star and $s=x^0-y^0$.  The $v$ integral can be performed analytically.  Note that Eq. (\ref{Pe}) and (\ref{X}) were derived for detectors in Minkowski spacetime, where $x^0(\tau)=y^0(\tau)=\tau$, but since we are only interested at the first order perturbation, the effect of the time dilation in the corrected propagator would be a second order term which we neglect.  We can thus use Eq. (\ref{Pe}) and (\ref{X}) with the first order correction of the propagator and with no time dilation, that is $x^0(\tau)=y^0(\tau)=\tau$. For the same reason, we can also use at this order $L_p=L$.  Therefore, we can put Eq. (\ref{G1}) in Eq. (\ref{Pe}) and we obtain the first order correction to the local noise $\delta P_{j(1)}$:
\begin{eqnarray}
\delta P_{j(1)}&=&\frac{\alpha^2\sigma \sqrt{\pi}}{\pi^2 r_j} \Re \Bigg\{ \int_0^{\infty}ds e^{-s^2/(4\sigma^2)-i\Delta E s} \int_0^{\infty} dR  \nonumber\\
&&\times  R\bar{h}^{00}(R)  \Bigg[\Bigg(\ln\left(\frac{2(r_j+R)+(s-i\epsilon)}{2(r_j+R)-(s-i\epsilon)}\right)\nonumber\\
&&-\ln\left(\frac{2|r_j-R|+(s-i\epsilon)}{2|r_j-R|-(s-i\epsilon)}\right) \Bigg)\frac{1}{4(s-i\epsilon)^3}\nonumber\\
&&-\frac{2(r_j+R)(2(r_j+R)^2-s^2)}{(s-i\epsilon)^2\left(4(r_j+R)^2-(s-i\epsilon)^2\right)^2}\nonumber\\
&&+\frac{2|r_j-R|(2(r_j-R)^2-s^2)}{(s-i\epsilon)^2\left(4(r_j-R)^2-(s-i\epsilon)^2\right)^2} \Bigg] \Bigg\}.\label{P1G}
\end{eqnarray}
Similarly for the exchange term $\delta X_{(1)}$, we put Eq. (\ref{prop1}) in Eq. (\ref{X}) and we then use a simple change of variables to obtain
\begin{eqnarray}
\delta X_{(1)}&=&-\frac{\alpha^2\sigma\sqrt{\pi}e^{-\sigma^2\Delta E^2}}{2\pi^2 r_1} \int_0^{\infty}ds e^{-s^2/(4\sigma^2)} \nonumber\\
&&\times \int_0^{\infty} dR R\bar{h}^{00}(R)\int_{|r_1-R|}^{r_1+R}dv_1 v_1\nonumber\\
&&\times\left[\frac{3(s^2+v_1v_2)(v_1+v_2)+v_1^3+v_2^3}{(v_1v_2+i\epsilon)(s^2-[v_1+v_2+i\epsilon]^2)^3}  \right]
\end{eqnarray}
where $v_2=\sqrt{v_1^2\left(1+\frac{L_p}{r_1}\right) +L_p\left(r_1+L_p-\frac{R^2}{r_1} \right)}$.  The $s$ integration can be performed analytically, such that we are left with a relatively simple expression for $\delta X_{(1)}$ which involves only two integrations:
\begin{eqnarray}
\delta X_{(1)}&=&\frac{\alpha^2\sigma\sqrt{\pi}e^{-\sigma^2\Delta E^2}}{2\pi^2 r_1} \int_0^{\infty} dR R \bar{h}^{00}(R) \int_{|r_1-R|}^{r_1+R}dv_1 \nonumber\\
&&\times\frac{1}{16 (v_2+i\epsilon)\sigma^4}\Bigg\{i\pi \mathrm{erfc}\left(\frac{-i(v_1+v_2)}{2\sigma} \right)\nonumber\\
&&\times e^{-(v_1+v_2))^2/(4\sigma^2)}\left[2\sigma^2-(v_1+v_2)^2\right]\nonumber\\
&&-2\sqrt{\pi}\sigma(v_1+v_2) \Bigg\}.\label{X1G}
\end{eqnarray}

\section{Negativity on the perturbed background}\label{negativity}

We now have all the tools to compute explicitly the corrected negativity. Using the $\bar{h}^{00}(|\vec{r}|)$ of Eq. (\ref{hb}) in Eq. (\ref{P1G}) and (\ref{X1G}), we can find $\delta P_{j(1)}$ and $\delta X_{(1)}$  by numerically evaluating the remaining integrals. $\tilde{P}_{j(1)}$ and $\tilde{X}_{(1)}$ can then be evaluated exactly using Eq. (\ref{TP}) and (\ref{TX}) such we end up with the full noise term $P_{j}=P_{(0)}+\tilde{P}_{j(1)}+\delta P_{j(1)}$ and the full exchange term $X=X_{(0)}+\tilde{X}_{(1)}+\delta X_{(1)}$ using Eq. (\ref{PG}) and (\ref{XG}) for the zeroth order terms.  This allows us to compute the negativity between the two detectors using Eq. (\ref{N}).

\begin{figure}
\centering
\includegraphics[scale=0.26]{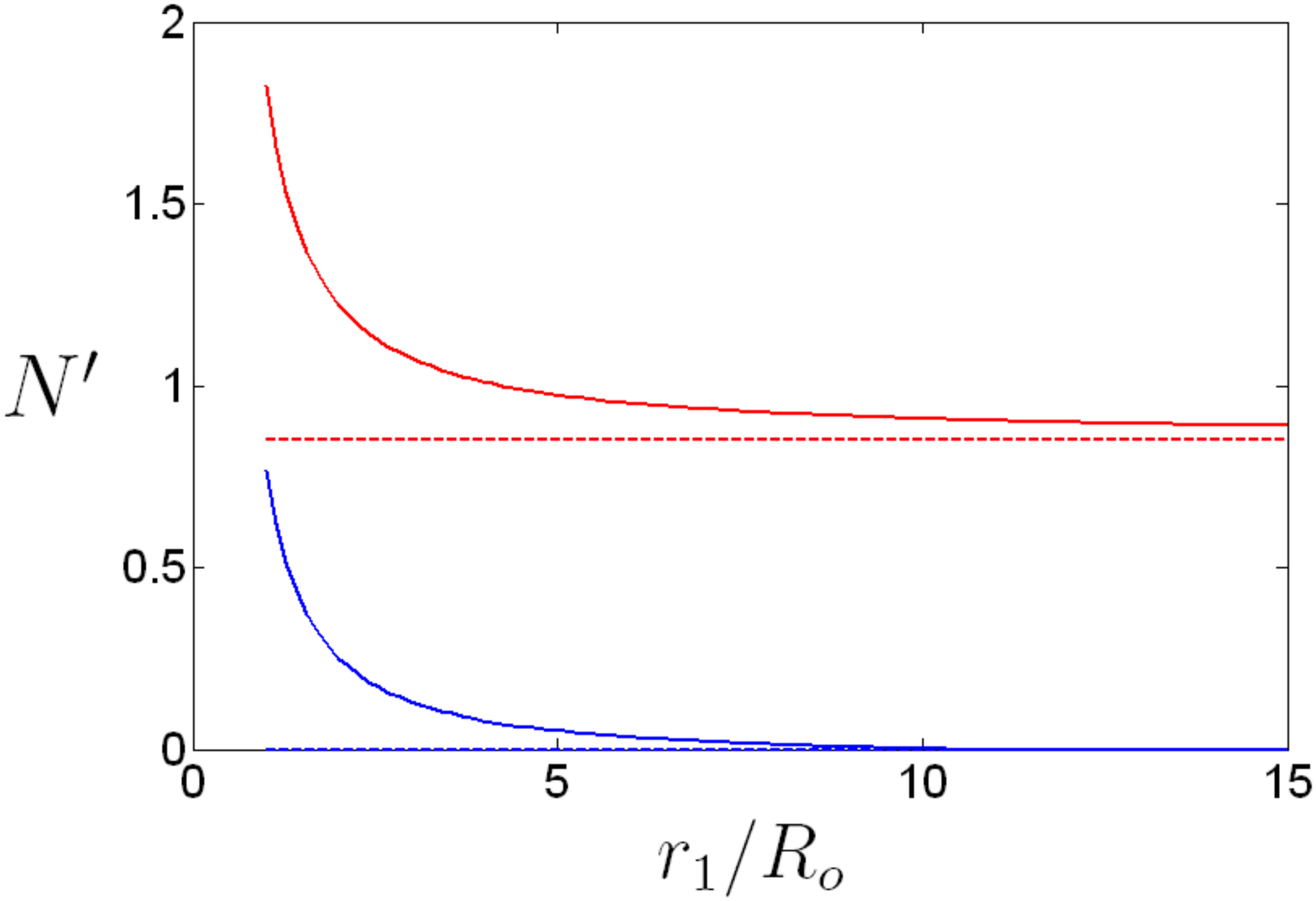}
\renewcommand{\figurename}{FIG.}
\caption{$N'=8\pi^2N/\alpha^2$ as a function of $r_1/R_o$. Here $\sigma\Delta E=0.00674$, $\Delta E R_o =1$ and $M/R_o=0.001$. The upper (red) curve correspond to  $L_p/R_o=0.0095$ and the lower (blue) curve to  $L_p/R_o=0.01$.  The upper and lower dashed lines are the asymptotes $r_1/R_o\rightarrow\infty$.}
\label{fig:NG3}
\end{figure}
Numerical evaluations indicate that $|X|$ linearly increases with the strength of the gravitational potential $M/R_o$ of the star while $P_j$ linearly decreases with $M/R_o$.  Therefore, the negativity $N$ linearly increases with the strength of the gravitational field $M/R_o$.  In a similar fashion, numerical evaluation of $|X|$ and $P_j$ indicate that the correction to the negativity $N$ decreases roughly like $R_o/r_1$ as $r_1/R_o\rightarrow\infty$ but remains positive, see Fig. (\ref{fig:NG3}).  On Fig. (\ref{fig:NG3}) we chose two sets of parameters $L_p/\sigma$ and $\sigma\Delta E$.  With the first one, the negativity $N$ is non-zero even without the gravitational field such that we only have entanglement enhancement by gravity.  With the second set of parameters, we have $N=0$ and $|X|\approx P$ without the gravitational field such that in that particular case we not only have entanglement enhancement by gravity but also entanglement creation by gravity.

We may heuristically interpret this phenomenon by looking at the local noise term $P_j$ and the exchange term $|X|$ separately.  Since the gravitational field increases the momentum of virtual particles near the star (as seen by the fixed detector $j$), it is  more energetically expensive to have many of them so the local noise has to decrease. As for the exchange term, we hypothesise that it increases because the gravitational field creates a lensing effect such that more virtual particles emitted by detector 2 hit detector 1.

As we previously mentioned this effect scales linearly with the strength of the gravitational field $M/r_1$, so for detectors with $\sigma\gg 1/\Delta E$
and $\sigma \gtrsim L_p$ we have for the Earth  $N_{(1)}\lesssim  10^{-9} N_{(0)}$ while for the Sun we have $N_{(1)}\lesssim 10^{-6}  N_{(0)}$.  Since entanglement swapping from the entanglement $N_{(0)}$ of the vacuum has still not been observed, we expect that observing $N_{(1)}$ will be very difficult.  Nevertheless, it should be interesting to see if this effect can be modeled in a quantum field analog like a linear ion trap \cite{Meni2}.

\section{Outlook}\label{outlook}

Our calculations depended on the assumption that the vacuum of the quantum field is described by the Boulware vacuum.  If we had considered two Unruh-DeWitt detectors near a black hole in an Unruh or a Kruskal vacuum \cite{Bir}, the Hawking temperature seen by both detectors would have increased the local noise significantly such that the entanglement between both detectors should be degraded, not enhanced. It should be interesting to investigate in detail to what extent entanglement extraction by detectors near black holes and stars is affected by the properties of the corresponding vacuum states.

In this context, it should also be interesting to investigate whether one can effectively model a black hole by using a confining potential, say on a shell.  Indeed, a trapping potential can have horizons, so it may be possible to have a non-trivial vacuum in which particle production occurs because of the potential.  Such an analysis could show the Hawking effect and its various open questions in a new light.

Since we observed that the exchange term $|X|$ increases because of the gravitational field, it is tempting to speculate on the Casimir-Polder force near a constant density star.  Indeed, the exchange term and the Casimir-Polder force have essentially the same interpretation, that is they are the result of a continuous exchange of virtual particles.  We therefore conjecture that Casimir or Casimir-Polder forces can slightly increase in a weak gravitational field.

\section*{Acknowledgments}

M.C. acknowledges support from the NSERC PGS program. A.K. acknowledges support from CFI, OIT, the Discovery and Canada Research Chair programs of NSERC.

\end{document}